# Dynamic Programming Algorithms for Discovery of Antibiotic Resistance in Microbial Genomes

*Manal Helal[1,2], Vitali Sintchenko[1,3]*

[1]*Centre for Infectious Diseases and Microbiology, Sydney West Area Health Service, NSW, Australia*

[2]*School of Computer Science, University of New South Wales, NSW, Australia*

[3]*Centre for Health Informatics, University of New South Wales, NSW, Australia*

*Abstract*

*The translation of comparative genomics into clinical decision support tools often depends on the quality of sequence alignments. However, currently used methods of multiple sequence alignments suffer from significant biases and problems with aligning diverged sequences. The objective of this study was to develop and test a new multiple sequence alignment (MSA) algorithm suitable for the high-throughput comparative analysis of different microbial genomes. This algorithm employs an innovative tensor indexing method for partitioning the dynamic programming hyper-cube space for parallel processing. We have used the clinically relevant task of identifying regions that determine resistance to antibiotics to test the new algorithm and to compare its performance with existing MSA methods. The new method "mmDst" performed better than existing MSA algorithms for more divergent sequences because it employs a simultaneous alignment scoring recurrence, which effectively approximated the score for edge missing cell scores that fall outside the scoring region.*

## 1. Introduction

The emerging genome sequencing technologies and bioinformatics provide new opportunities for studying life-threatening human pathogens and to develop innovative decision support tools for the diagnosis and treatment of infections. The accumulation of sequenced genomes of bacteria showed a good fit to exponential functions with a doubling time of approximately 20 months, however, their high-quality comparative genomic analyses require adequate methods and tools [1]. Closing the gap between our emerging capacity to generate vast quantities of sequencing data and our ability to ensure high quality analyses will remain the major goal of the next decade. Thus the infectious disease informatics leads to more targeted and effective approaches for prevention, diagnosis and treatment of infections through a comprehensive review of the genetic repertoire and metabolic profiles of a pathogen.

Clinical genomics and informatics have been dominated by human genome paradigms in which genomic rearrangements typically denote dysfunction. However, microbial genomes, particularly those of bacteria, have a mosaic structure and may vary significantly, even within a species; it remains unclear how microbial genomic data should be processed so that it is easily interpretable, accessible and sharable. The great diversity of mutational patterns contributing to antimicrobial resistance complicates the choice of optimal therapies. A range of informatics tools, to predict drug resistance or response to therapy from genotype, have been developed to provide clinician support [2,3]. These tools use either a statistical approach, in which the inferred model and prediction are treated as regression problems, or machine learning algorithms, in which the model is addressed as a classification problem [2,3]. Many new molecular-based technologies (proteomics, transcriptional profiling, study of gene expression *in vivo*) have originated or have expanded into wider use, and have been made possible by the availability of complete bacterial genome sequence information and subsequent informatics tools. Taken together, these technologies, overlaid within an established drug discovery program, now afford the opportunity for the identification, validation, and process design for high-throughput target mining [4].

Microbial genomes are thousands or millions of base pairs in length and their analysis demands efficient techniques of multiple sequence alignment (MSA). Existing methods have been developed for short sequences and cope well with relatively similar sequences; otherwise, higher likelihood of errors in the alignment produced is expected. The objective of this study was to test a new MSA algorithm and to analyse its suitability for the high-throughput comparative analysis of microbial genomes.

## 2. Motivation – limitations of existing MSA methods

An accurate MSA is critical for answering different research, clinical and diagnostic questions in biomedicine. Applications include clustering of sequences, phylogenetic tree reconstruction, secondary or tertiary structure prediction, function prediction, polymerase chain reaction (PCR) primer design and data validation. These applications are affected by the main limitation of the existing MSA tools, which is the inability to align divergent sequences. This work presents an MSA tool that is capable of constructing meaningful MSA of divergent sequences. This property enables the identification of functional regions in sequences responsible of a specific behaviour, by aligning sequences of known opposite functions.

The relevance of a mathematical optimal solution to a biological/clinical meaningful one is still an active area of research. MSA belongs to the hard NP-complete class of problems that is believed that its solution requires super polynomial computation efforts, and a solution for it can solve all problems in its class as proven in [5] and [6]. This means that an optimal solution cannot be obtained using a reasonable amount of computation time. To solve large NP-complete problems, one has to balance the optimality requiring large, possibly intractable computation time with faster but often sub-optimal solutions. The first approach presents optimization problems that can be solved by enumeration methods such as the cutting plane, branch and bound, integer, linear and non-linear programming and dynamic programming [7]. The second approach relies on approximation (heuristic) algorithms, like local search and randomization algorithms. Some methods can be configured to be an optimization or an approximation algorithm, for example, by introducing heuristic bounding rules to a branch and bound algorithm.

Existing methods are mostly progressive and iterative. The progressive methods compare sequences two by two (pair-wise) first to build a distance matrix. Then from the distance matrix, it uses a clustering algorithm such as UPGMA [8] or NJ [9] to generate a rooted binary guide tree that is traversed to construct the final MSA. The latter falls in local optima in the early stages and cannot be corrected later. Iterative methods (e.g., MUSCLE [10] and PRRP [11]) attempt to correct this weakness by involving two nested iterative loops that optimize the alignment with respect to a guide tree in the inner loop, and re-estimate the guide tree by using the current MSA in the outer loop. However, early errors in the alignment can still propagate to the final MSA. The only way to avoid these mistakes is by simultaneously using all information contained in the sequences; however, it is computationally prohibitive [7].

Consistency-based methods were originally described by GOTOH, followed by an exact approach described by Kececioglu in a linear programming formulation of MSA using a Maximum Weight Tree (MWT) [12]. A heuristics developed by Morgenstern expanded the applicability of the MWT method using overlapping weights and define consistency to describe the compatibility of a pair of matched segments to the partially defined MSA [13]. The main idea behind consistency based methods was to employ a third sequence in the evaluation of any pair of sequences to increase the consistency all over the alignment. TCoffee combined the progressive approaches with the consistency based approach [14]. So far, M-Coffee remains the only package that allows multiple MSAs using any selected method. M-Coffee turns them into a library which is used to construct an MSA that is consistent with the original alignments. The resulting alignment is usually slightly more accurate than ones produced by any of individual alignment methods. This observation suggests that the current MSA aligners have reached saturation point and can not be further improved without additional information about biological functions. This information includes functional data in the form of transcript structure, structural data (e.g., targetDB) and protein/DNA interactions (e.g., ChIp-Chip data). The template-based alignment has been proposed to map extra available information to the sequence information to guide the alignment. This extra information can be a 3D structure, a profile, or any kind of the structure-function prediction. A structural extension is provided through predicted structural templates such as PDB homolog for proteins (3D coffee – EXPRESSO and PROMAL 3D packages do a BLAST against PDB database to retrieve this template) or RNA secondary structure (T-Lara, MARNA and R-Coffee packages). Homolog extensions, on the other hand, use profile templates by replacing sequences with profiles of homologs that is built by methods like PSI-Blast. Homolog templates are utilised in MSA tools such as PRALINE, PROMALS and TCoffee (version 6+ and psicoffee mode). Evidence suggests that template based MSA, especially the ones with structural extensions can be more accurate than other methods that do not consider extra information, or consider homology extensions. However, the over-reliance on templates can bias the alignment to known knowledge [7].

Simultaneous alignments do not suffer from problems with local optima or bias. These methods are more robust against parameter changes. The quality of alignment increases as more members of the same family are aligned with sequences from outside the family [15]. However, these techniques remain computationally expensive. One of the successful techniques used to decrease the complexity of the dynamic programming for multiple sequence alignment is the "sum-of-pairs" algorithm. It reduces the computation steps to the alignment of every pair of sequences (on the surfaces of the scoring hyper-cube), and then sums the alignment scores to align all the sequences on an internal diagonal that crosses all the dimensions producing the total alignment. This is known as the Carrillo and Lipman bounds, which have been observed to be overestimated [16]. The limitation of this algorithm is that it calculates a weighted sum of its projected pair-wise alignments. Simultaneous alignment tools such MSA [17], DCA [18], and linear programming formulation of MSA [12] can fit biological intuition more closely. Simultaneous methods either approximate the higher dimensional space into two-dimensional spaces as shown in the Carrillo and Lipman approaches, or require a high dimensional indexing. For sequence searching (not MSA construction but can be exploited to form MSA), indexing methods can be used such as the B+-tree indexing system [19]. Other indexing structures, such as R*-trees, X-trees and SR-trees, perform well only with short sequences. Hash table based sequence search methods include FASTA, and BLAST work by constructing a hash-table on one sequence and insert all substrings of length l. Then the tool finds all exactly matching substrings (seeds), and extends in both directions of each seed, and combines them to find better alignments. Suffix trees methods have been proposed to handle similar sequence search problems. However, they handle mismatches inefficiently and consume a lot of memory space. Vector space indexing methods, such as VP-tree, partition the data space into spherical cuts by selecting random reference points from the data. MVP-tree is a variation of VP-tree that uses more than one point at each level. Unfortunately, vector-space methods remain inefficient for highly dimensional data.

Embedding-based indexing techniques like FastMap and MetricMap can be applied successfully only to low dimensional (less number of sequences) databases, because of the time consuming process of finding a mapping function that preserves the distances. Height balanced tree such as M-tree and Slim-tree (as an evolved M-tree that reduced the amount of overlapping nodes) attempt to reduce the height of the tree at the expense of flexibility in reducing the overlapping nodes. DBM-tree is an unbalanced tree that reduces the overlapping nodes in high-density regions. The work in [20] proposed a novel index structure, SEM-tree (Sequence Embedding Multiset tree) based on the Sequence Dimensionality Reduction (SDR) method. The SEM-tree levels represent a compression level with increasing length of the multiset towards the leaves level where the original sequences are stored. This compressed representation of the sequences makes the sequence comparison which is based on the number of distinct characters that form the sequence set, instead of the length of the sequences. The recursive traversal of the tree can be predict whether the sub-tree is included in the final result or not and hence decide whether to continue or stop the recursion. The approximation of the compressed representation can be bounded to avoid false negatives.

In all these methods care must be taken in choosing scoring matrices and gap penalties. To summarize the problems of existing methods:

- Most of them assume minimum percent identity of approximately 40% for proteins and approximately 70% for DNA [21, 22]. The resulting alignments are acceptable for families of moderately diverged sequences. Otherwise, a much higher likelihood of errors in the alignment will be expected since these methods can easily run into local optima like any hill-climbing bottom-up methods.

- The discussed global methods require the sequences to be related over their whole length or at least most of it, with the exception of DiAlign [23]. In addition, these methods are sensitive to the order in which the sequences are input. This is due to the fact that they calculate a guide tree based on that order.

- Additional methods like PRRP [11], Hidden Markov Models (HMMs) [24,25,26] and Simulated Annealing [27] have been applied to refine an MSA produced by another method.

- In addition, these methods are sensitive to the order in which the sequences are input. A different ordering of sequences will produce different alignments. This is due to the fact that they calculate a guide tree based on that order.

- Progressive methods depend on pair-wise alignments, which is less sensitive than simultaneous alignment. This is because pairs of already aligned subfamilies (or closely related sequences) are calculated first, and there is usually more than one optimal alignment of the pairs and the choice of one of them might not be the optimal for the other pair-wise alignments nor have the highest biological relevance.

- The pair-wise alignments are dependent on the parameters used in the calculations and the parameter changes will not be reflected in the resulting MSA optimal alignment. The pair-wise alignments can cause bias in the positioning of gaps [21] and statistical uncertainty in the produced conclusions [22].

- Template-based MSA methods assume some knowledge about the outcome (e.g. using annotations and biological knowledge in the objective function, or adding homologs or profiles to the sequences dataset). These methods also presume conserved order of aligned residues, with the exception of ABA [28], ProDA [29], TBA [30], and MAUVE [31].

The advances in parallel and distributed processing enabled research into parallelising large and high dimensional problems. The main requirement for parallel processing is data partitioning and dependency modelling. Performance can be optimised by employing more processing elements while reducing communication between them. Attempts to parallelize sequence comparisons have been so far limited to pair-wise alignments [32]. Simultaneous multiple sequence alignment methods can rely on high dimensional indexing techniques where overlaps make its partitioning a non-uniform task.

## 3. Definitions and Methodology

### 3.1. Alignment of antibiotic resistance determining regions

Regions of similarity or dissimilarity between a set of sequences, obtained from pathogens with known resistance or susceptibility to antibiotics (class of quinolones in our case) were explored. Mechanisms of resistance to quinolones have been studied extensively in many different bacterial species and are usually due to single point mutations in the target of these drugs, DNA gyrase. Resistance mutations most often occur within a stretch of 50 nucleotides, the so called "quinolone-resistance determining regions" (QRDRs), which are located in the genes for the A subunits of the enzyme *gyrA* gene [33]. The resistance mutations in *gyrA* codons 84 or 88 usually lead to the high-level *in vitro* resistance but other mutations can also infrequently occur.

### 3.2. Sources of data

The set of *gyrA* gene sequences were extracted from following microbial genomic data available in the GenBank: *Mycobacterium tuberculosis* (NCBI Accession Number NC_000962, sequence length 2518bp); *Mycobacterium kansasii* (NCBI Accession Number Z_68207, sequence length 1648bp); *Staphylococcus aureus* MSSA476 (NCBI Accession Number NC_002953, sequence length 2665bp); *Mycoplasma pneumoniae* (NCBI Accession Number NC_000912, sequence length 2443bp); *Clostridium difficile* (NCBI Accession Number NC_009089, sequence length 2521b); and *Treponema pallidum* (NCBI Accession Number NC_010741, sequence length 2428bp). Gene sequences of *M.tuberculosis, M. kansasii* and *Staphylococcus aureus* MSSA476 were grouped as quinolone susceptible. Gene sequences of *Treponema pallidum, Clostridium difficile* and *Mycoplasma pneumoniae* were classified as resistant to quinolone thus potentially harbouring changes in the *gyrA* gene.

### 3.3. MSA algorithms

Based on the multidimensional optimal dynamic programming algorithm as a simultaneous alignment technique, an innovative tensor (high dimensional space or hyper-cube) indexing scheme has been used as the foundation of the MSA tool used in this study. This tensor indexing scheme has enabled a uniform partitioning of the scoring hyper-cube space to parallel processing elements as described in [34] and [35]. The partitioning of the scoring tensor keeps the communication to the minimum, by clustering the assignment to processors based on neighbourhood of partition indices. The resulting massively parallel solution can employ as many processors as the input dataset require. A further search space reduction technique was developed to reduce the scoring space to the area where an optimal alignment is expected (around the hyper-diagonal of the scoring space) [36]. While all methods in the literature uses the Carrillo and Lipman bounds to decide how much to cut from the edges of the reflections on the surface of the scoring hyper-cube to reduce the search space, this new technique decides how far from the hyper-diagonal we need to score to reach an optimal alignment. The solution runs on computer clusters, multi-core architectures, and high performance machines, due to the usage of Message Passing Interfaces (MPI) that hide the hardware architecture details. The new MSA method is called "mmDst".

The MSA is organised in the following steps:
1. Multiple sequence alignment of the first set of sequences (sensitive to antibiotics) to derive a consensus sequence, or a profile of the known behaviour.
2. Align the sensitive consensus to the highest resistant sequence "*Treponema pallidum*" to identify major differences.
3. Align a set of resisting sequences and derive their consensus sequence,
4. Align the consensus of the sensitive sequences to the consensus of the resisting sequences to identify the regions of similarity and dissimilarity (visually or calculated from the scores) of both profiles.

The mmDst method was compared to existing MSA *heuristic* methods such as CLUSTAL W [37], MUSCLE [10], TCoffee [14], Kalign [38] and MAFFT [39]. As mentioned in the introduction section, these methods are based on pair-wise alignments, which are proven to be less sensitive than simultaneous alignments [15]. The web portal of EMBL-EBI for different MSA methods was used [40] to compare and evaluate the results. They all relied on the identity matrix for scoring DNA sequences. Default parameters were mostly used in all methods, except where there was an interface to make them score as similar as possible. CLUSTAL W used gap opening penalty = *15* and gap extension penalty = 6.66. MUSCLE used gap opening penalty = 15 and gap extension = 1. MAFFT used gap opening = 1.53 and an extension gap penalty = 0.123.

### 3.4. mplementation

The mmDst method was tested on small HPC machines and one SGI Altix cluster of maximum 64 nodes. The processor scalability reduces the execution time as more processors were employed to achieve the minimal communication cost, and high data locality. The system was implemented on a SunFire X2200 with 2xAMD Opteron quad processors of 2.3 GHz, 512 Kb L2 cache and 2 MB L3 cache on each processor, and 8GB RAM. The sequences were aligned on a reduced search space factor "Epsilon" equals 1, which represented 0.21% of the search space for the sensitive sequences and 0.19% of the resisting sequences. The pair-wise alignments of the consensus sequences were done in full search space. The score of one cell in the hyper-plane was based on the maximum values of the $2^k-1$ neighbours' temporary scores. The latter was calculated as the total pair wise scores of all its corresponding residues on all dimensions (sequences) corresponding to a decremented index element from the current cell index to the neighbour index, plus multiplication of the gap score by the number of un-decremented index elements. The following penalties were applied: gap opening = -4, gap extension = -2, mismatch score = -1, and match score = 1.

The Sum of Pairs Score is usually used to assess the performance of MSA methods. This score increases as the program succeeds in aligning more matching residues in each column in the final alignment, with minimum gap insertions all over, assuming statistical independence between columns [41]. Shannon entropy is a simple quantitative measure of uncertainty in a data set. In the context of drug resistance as conferred from single mutations, knowledge of the frequencies of different amino acids in the mutation position as drawn from resistant and sensitive populations, will enable us to guess the amino acids responsible for the resistance. This is because these amino acids were certain (low entropy) in the sensitive population, versus the uncertain (high entropy) in the population with high level resistance to quinolones [42]. To identify the exact start and end of the regions of highest and lowest column scores in the alignment, a simple method was implemented. The alignment was scanned for all regions of width = 2 * the window size used in the plots of the results section. Using the sum of pairs scores generated in 1, every region was given a score using the following average function:

*Average Region score = sum($c_i$) / (Window Size)*

Where $c_i$ is the column sum of pairs score using the identity matrix for each column *i* within the region.

## 4. Results

The mmDst algorithm successfully identified QRDRs and handled sequences of different length. The quality of the alignments produced by the different MSA tools is assessed by the sum-of-pair score of the alignment (Table 1) and the entropy of the information retained in the alignment (Table 2). Both Table 1 and 2 measure the alignment quality scores for the different alignments done in this experiment. The columns in both tables measure the score for the alignment of the quinolones sensitive sequences alone, quinolones resistant sequences alone, the set of quinolones sensitive sequences and "Treponema pallidum" as the most quinolones resistant sequence, and the alignment of the consensus sequence of the first alignment and the second alignment respectively. The most accurate (highest scoring alignment) was achieved in the third column in Table 1 and 2, i.e. the alignment of the set of quinolones sensitive sequences and "Treponema pallidum". The highest and lowest region score were determined ( Table 2: Entropy value for the aligments produced by the different methods for the Sensitive Sequences Alignment, Resistant Sequences Alignment, Sensitive Sequences Consensus Alignment with the most resistant sequence of Treponema pallidum, and Sensitive Sequences Consensus and Resistant Sequences Consensus Alignment.3) for the alignments of the sensitive consensus sequence with "Treponema pallidum", and for the alignments of the antibiotic sensitive consensus sequence with resistant consensus sequence. The *gyrA* gene of intrinsically quinolone-resistant *Treponema pallidum* demonstrated significant dissimilarity from *gyrA* genes sequences obtained from quinolone susceptible organisms of Mycobacteria (Figure 1).

|  | Sensitive Seq | Resistant Seq | Sen & TP | Sen & Res Cons |
|---|---|---|---|---|
| mmDst | 339 | 2231 | 849 | 1582 |
| MUSCLE | 439 | 3216 | 640 | 1123 |
| TCoffee | 443 | 2881 | 520 | 1025 |
| CLUSTAL W | 222 | 1966 | 478 | 1469 |
| Kalign | -1593 | -716 | -285 | 1389 |
| MAFFT | -3647 | -4712 | -1670 | -2114 |

*Table 1: Sum-of-Pairs Scores for the alignments produced by the different methods for the Sensitive Sequences Alignment, Resistant Sequences Alignment, Sensitive Sequences Consensus Alignment with the most resisting sequence "Treponema pallidum", and Sensitive Sequences Consensus and Resistant Sequences Consensus Alignment.*

|  | Sensitive Sequences | Resistant Sequences | Sensitive sequences & sequence of Treponema pallidum | Sensitive & Resistant Consensus Alignment |
|---|---|---|---|---|
| mmDst | 23869.74 | 27932.28 | 15362.36 | 15430.20 |
| MUSCLE | 26855.33 | 25144.80 | 16815.42 | 17264.06 |
| TCoffee | 27246.99 | 25682.06 | 17355.43 | 17797.34 |
| CLUSTAL W | 28362.00 | 28240.50 | 17753.33 | 18836.40 |
| Kalign | 33336.24 | 34849.35 | 21156.91 | 22011.64 |
| MAFFT | 37834.49 | 40707.34 | 26597.46 | 37938.70 |

*Table 2: Entropy value for the aligments produced by the different methods for the Sensitive Sequences Alignment, Resistant Sequences Alignment, Sensitive Sequences Consensus Alignment with the most resistant sequence of Treponema pallidum, and Sensitive Sequences Consensus and Resistant Sequences Consensus Alignment.*

|  |  | Sensitive consensus sequence & the "Treponema pallidum" Alignment | | | Sensitive consensus sequence & the resisting consensus sequence alignment | | |
| --- | --- | --- | --- | --- | --- | --- | --- |
|  |  | Score | From | To | Score | From | To |
| mmDst | Highest | 0.64 | 151 | 351 | 0.80 | 272 | 472 |
| mmDst | Lowest | 0.07 | 2167 | 2367 | 0.33 | 1567 | 1767 |
| MUSCLE | Highest | 1.03 | 450 | 650 | 1.11 | 356 | 556 |
| MUSCLE | Lowest | -0.83 | 2375 | 2575 | -0.89 | 2668 | 2868 |
| TCoffee | Highest | 1.06 | 233 | 433 | 1.15 | 430 | 630 |
| TCoffee | Lowest | -0.98 | 2495 | 2695 | -0.36 | 2737 | 2937 |
| CLUSTAL W | Highest | 0.98 | 233 | 433 | 1.05 | 292 | 492 |
| CLUSTAL W | Lowest | -0.25 | 2176 | 2376 | 0.18 | 2589 | 2789 |
| Kalign | Highest | 0.54 | 135 | 335 | 0.90 | 3265 | 3465 |
| Kalign | Lowest | -0.54 | 3253 | 3453 | -0.17 | 0 | 200 |
| MAFFT | Highest | 0.52 | 101 | 301 | 0.58 | 3915 | 4115 |
| MAFFT | Lowest | -1.72 | 2654 | 2854 | -2.00 | 427 | 627 |

*Table 3: Highest and Lowest (maximum and minimum Sum-of-Pairs scores respectively) Regions (as identified in the "From" base pair number "To" base pair number) of Similarity or Dissimilarity in the alignment of the sensitive sequences consensus sequence and the "Treponema pallidum" sequence as per alignment method in the left hand side columns, and the sensitive sequences consensus sequence and resisting sequences consensus sequence alignment.*

The similarity regions plots shown in Figure 1: Similarity Regions Plot (averaged on 100 bp on the x-axis as relative residues positions) of the alignment (measured by the SP score on the y-axis) of the consensus sequence of the sensitive sequences with the most resisting sequence "Treponema pallidum" using the six different methods: a) mmDst, b) MUSCLE, c) TCoffee, d) CLUSTAL W, e) Kalign, f) MAFFT.1 are generated by plotcon algorithm averaged on a window size of 100 base pairs. The difference alignment methods used show different areas of similarity (regions where the y-axis score is higher) and dissimilarity (regions where the y-axis score is lower), according to the SP score of the columns corresponding to the 100 base pairs averaged on the x-axis.

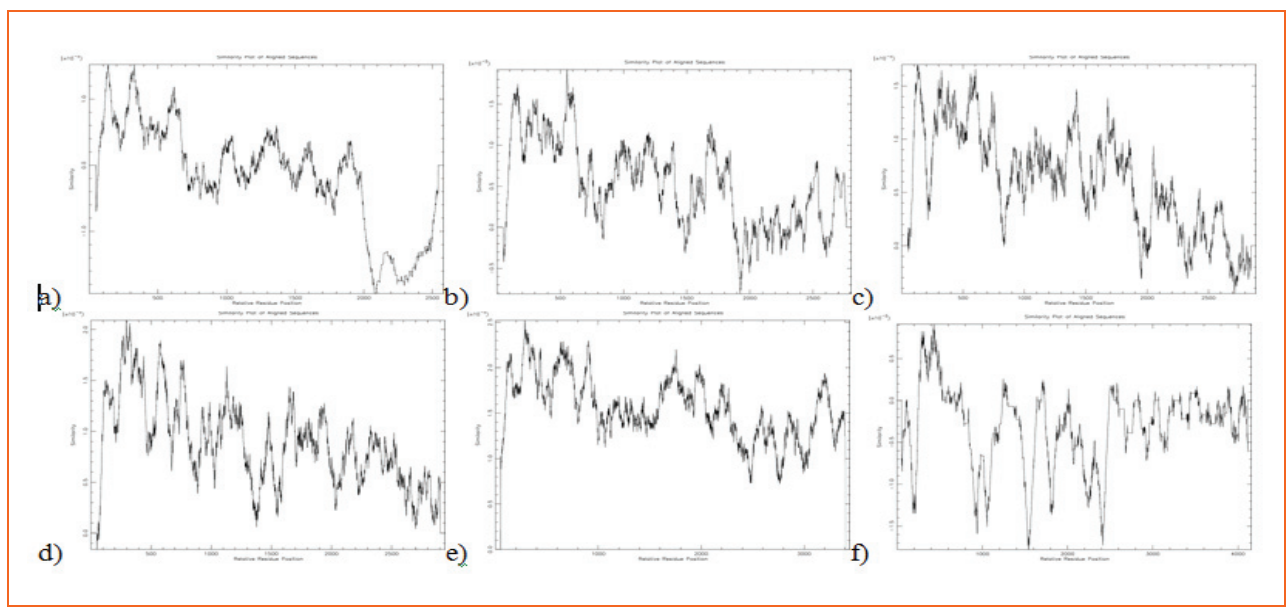

*Figure 1:* Similarity Regions Plot (averaged on 100 bp on the x-axis as relative residues positions) of the alignment (measured by the SP score on the y-axis) of the consensus sequence of the sensitive sequences with the most resisting sequence "Treponema pallidum" using the six different methods: a) mmDst, b) MUSCLE, c) TCoffee, d) CLUSTAL W, e) Kalign, f) MAFFT.

Table 1 and 2 show that the proposed method "mmDst" score came third after TCoffee and MUSCLE in the first two cases (columns 1 and 2), where similar sequences were aligned. MUSCLE, TCoffee and CLUSTAL W are progressive methods based on pair-wise alignments and building a guide tree based on an objective function. These methods work well with sequences of assumed similarity of 90% or higher.

However, in the third alignment case (third column in Table 1 and 2) where the consensus sequence of the alignment of the quinolones sensitive sequences were aligned with the most antibiotic resistant sequence which is "*Treponema pallidum*", mmDst score came second after MUSCLE. In the fourth case (fourth column in Table 1 and 2), which is the alignment of the consensus sequence of the set of sensitive sequences with the consensus sequence of the set of resisting sequences, mmDst scored the highest over all other methods. These findings demonstrate that mmDst scores better when aligning sequences of large dissimilarity and can identify highly dissimilar regions along the full length of the input sequences.

## 5. Discussion

The direct comparison of six MSA algorithms highlighted significant challenges in comparative genomics of pathogens. The majority of high-quality algorithms are computationally expensive to be implemented in routine diagnostic laboratories. Furthermore, existing methods are sensitive to the order of sequence inputs order as a different ordering of sequences generates different alignments. Progressive methods rely on pair-wise alignments, which is less sensitive than simultaneous alignment. This phenomenon can be explained by the fact that pairs of already aligned subfamilies (or closely related sequences) are calculated first, and there is usually more than one optimal alignment of the pairs and the choice of one of them might be neither optimal for the other pair-wise alignments nor has the highest biological relevance. The pair-wise alignments are also dependent on the parameters used in the calculations and the parameters changes are not reflected in the resulting MSA optimal alignment. The pair-wise alignments can be biased because of the positioning of gaps [21] or statistical uncertainty [22].

Interestingly, all programs aligned better sequences of medium length and long sequences than short DNA sequences. The only exception was CLUSTAL W algorithm that improved traditional progressive methods. This phenomenon can be explained by the usage of an alternative Neighbouring-Joining algorithm for a guide tree construction, sequence weighting, as well as by position-specific gap-penalties. CLUSTAL W offers the choice of residue comparison matrix depending on the degree of identity of the sequences. MUSCLE aligned 5,000 sequences of average length 350 in 7 minutes on a desktop computer, requiring less time than all other tested methods, including MAFFT. MUSCLE and TCoffee produced, on average, the most accurate alignments, with 6% more positions correctly aligned than CLUSTAL W. It calculated the evolutionary distance between each pair of sequences. Then the method employs resulting distance matrix to cluster the sequences using UPGMA giving a binary tree. The tree is then used to construct a progressive alignment by aligning profiles of the two sub-trees at each internal node. TCoffee allowed the combination of a collection of multiple/pair-wise, global or local alignments into a single model. It is based on a 'greedy' progressive method that allows better use of information in the early stages, to rectify the problem with progressive methods of having errors happening early in the alignment and not being able to rectify it later. It also estimates the level of consistency of each position within the new alignment with the rest of the alignments. Kalign applied the same progressive method with the difference in the distance calculations which are based on the Wu-Manber approximate string-matching algorithm. MAFFT is a multiple sequence alignment based on Fast Fourier transform. It offers different levels of sensitivity. The new method "mmDst" scored better for more divergent sequences because it employs an innovative simultaneous alignment scoring recurrence.

Computer-assisted therapy is an attractive way to reduce the complexity of prescribing antimicrobial combinations [3]. It highlights the need for databases that can be widely shared, and allow correlation of quality-controlled data from genotypic resistance assays and treatment regimens with short- and long-term clinical outcomes. Differences in antimicrobial sensitivity reflect variation in amino acid composition of resistant microbes but, simply counting mutations enough to detect most functional differences, which affect treatment outcomes. The data linkages between laboratory and clinical databases will unlock the full utility of microbial profiles [2,3].

## 6. Conclusion

In summary, a growing amount of bacterial genomic data strengthened and streamlined the study of pathogens and offered new type of data for clinicians. This new paradigm of clinical data analysis has placed significant demands on the health informatics and bioinformatics support including the development of new algorithms for comparative genomics and dynamic programming to support high-throughput data handling in biomedicine. The new method "mmDst" performs better for more divergent sequences because it employs a simultaneous alignment scoring recurrence, which effectively approximated the score for edge missing cell scores that fall outside the scoring region [36]. With the newly added feature of search space reduction, mmDst can scale better with longer sequences. However, mmDst may not scale well with increased number of sequences, as heuristics methods do, because of the computational complexity of the high dimensional scoring recurrence. Further employment of parallel/distributed hardware architectures can be employed, by further reducing the memory consumption of "mmDst" and other less computationally demanding scoring functions, such as Sum-of-Pairs and viterbi.

The comparative experiments conducted in this study contrasted properties of MSA algorithms and highlighted their capacity for the rapid identification of genomic regions potentially responsible for the drug resistance. These methods may assist in the assessment of both mutation patterns and mutation frequency in clinically significant microbial genomes. Aligning the profiles of both families revealed a better visual identification of the similar and dissimilar regions, rather than the alignment of one sequence representative of one family to the consensus of the other. Alignment methods that are capable of automatically comparing diverged sequences can provide additional insights about genes responsible for specific clinical phenotypes.

## Acknowledgements

Support from The University of Sydney grant to use the SunFire HPC is gratefully acknowledged.

## Correspondence

Manal Helal
Centre for Infectious Diseases and Microbiology
University of Sydney
Sydney NSW, Australia
mhelal@usyd.edu.au